\documentclass{article}

\usepackage{arxiv}

\usepackage[utf8]{inputenc} 
\usepackage[T1]{fontenc}    
\usepackage{hyperref}       
\usepackage{url}            
\usepackage{booktabs}       
\usepackage{amsfonts}       
\usepackage{nicefrac}       
\usepackage{microtype}      
\usepackage{lipsum}
\usepackage{natbib}
\usepackage{graphicx}
\usepackage{verbatim}
\usepackage{color}
\usepackage{amsmath}
\usepackage{float}
\usepackage{tikz}
\usepackage{subfigure}

\graphicspath{ {./images/} }

\title{Modelling convective cell occurrence in proximity to cold fronts using extreme gradient boosting}

\author{
George Pacey \\
  Institute of Meteorology\\
  Freie Universität Berlin\\
  Berlin, Germany\\
  Current affiliation: University of Bern, Switzerland\\
  \texttt{george.pacey@unibe.ch}
  \And
 Stephan Pfahl \\
  Institute of Meteorology\\
  Freie Universität Berlin\\
  Berlin, Germany\\
  \texttt{stephan.pfahl@met.fu-berlin.de}
  \And
 Lisa Schielicke \\
  Department of Physics and Astronomy\\
  The University of Western Ontario,\\
  London, Canada\\
  \texttt{lschieli@uwo.ca}
}

\begin{document}
\maketitle
\begin{abstract}
Machine learning is emerging as a valuable tool for convection-related applications such as post-processing numerical weather prediction output, improving understanding of convective storm climatology and potentially improving existing convective parameterization schemes. In a rapidly developing field, it is vital to assess the strengths and limitations of machine learning approaches across different applications. Here, a probabilistic model is developed using a convective cell dataset as ground truth and predictors primarily from ERA5. The model’s ability to reproduce the convective cell climatology at different regions relative to cold fronts (i.e. post-frontal and pre-frontal) is assessed during the warm-season in Germany. The optimal number of features (predictors) is selected using a feature elimination strategy. Overall, the optimised model exhibits high skill in reproducing the spatial and temporal cell frequency at different regions relative to the front. While the highest cell frequency is correctly identified near the surface front, the model underestimates the actual cell count in this region. Feature importance analysis shows that the model depends most heavily on CAPE to make its predictions. Additionally, the time of day predictor is key for accurately capturing the diurnal cycle of convective cells on both sides of the cold front. The study highlights both the advantages and the limitations of data-driven models, offering valuable insights for future data-driven climate and weather prediction models. 
\end{abstract}


\section{Introduction}

The use of data-driven approaches has become increasingly popular in recent times. In the field of meteorology and climate, the interest has extended beyond meteorologists, also to technology companies such as NVIDIA, Google (Deep Mind) and Huawei, who have developed global data-driven weather forecasting models (\citeauthor{Pathak_et_al_2022}, \citeyear{Pathak_et_al_2022}; \citeauthor{lam_et_al_2023_graphcast}, \citeyear{lam_et_al_2023_graphcast}; \citeauthor{bi_et_al_2023_pangu}, \citeyear{bi_et_al_2023_pangu}; \citeauthor{Price_et_al_2024}, \citeyear{Price_et_al_2024}). ECWMF has also developed an artificial intelligence weather prediction model (AIFS; \citeauthor{lang_2024_et_al}, \citeyear{lang_2024_et_al}). The models can produce realistic thermodynamic and dynamic atmospheric fields forward in time, exhibiting on par or superior skill to those produced by conventional numerical weather prediction (NWP) (\citeauthor{ben_bouallègue_et_al_2024}, \citeyear{ben_bouallègue_et_al_2024}). Data-driven approaches rely on large amounts of historical data for training. During the training, the goal is to establish relationships between the predictors and the target variable(s). These relationships should also be generalisable outside of the training data. One main advantage compared to conventional NWP is that once the model is trained it is relatively inexpensive computationally to produce predictions. Most machine learning weather prediction (MLWP) approaches still utilise physics-based ERA5 reanalysis data to generate the training datasets, which of course involves using convectional NWP. Work on purely observational-based MLWP models is ongoing (e.g. \citeauthor{Allen_et_al_2025}, \citeyear{Allen_et_al_2025}; \citeauthor{ecmwf_ai_2025}, \citeyear{ecmwf_ai_2025}). \\

The aforementioned models have not yet directly addressed convective phenomena on the convective-scale, but there is some evidence that they can represent convective environments (e.g. wind shear and instability) on par with NWP models (\citeauthor{Feldmann_et_al_2024}, \citeyear{Feldmann_et_al_2024}). Atmospheric convection is associated with some of the most violent natural phenomena on earth such as lightning, hail, extreme precipitation and tornadoes. Strong events can lead to financial losses of the order 1 billion euros (e.g. \citeauthor{Kunz2018}, \citeyear{Kunz2018}; \citeauthor{Wilhelm2021}, \citeyear{Wilhelm2021}) and extreme convective precipitation events can lead to hundreds of fatalities (\citeauthor{ECHO_2024}, \citeyear{ECHO_2024}). Despite advancements in the representation of convection in numerical models, models are often still associated with biases and challenges remain (\citeauthor{Yano_et_al_2018}, \citeyear{Yano_et_al_2018}). Given the increased availability of observations, e.g. lightning detection networks, satellite imagers and crowdsourced severe weather reports, and the fundamental non-linear nature of convection, machine learning (ML) could prove to be a very useful tool to enhance understanding and prediction of convective phenomena.

There is already literature on applications of data-driven approaches for convective storm prediction, most of which is regional and often applied to post-process output from conventional NWP. Random forest uses an ensemble of decision trees to make predictions and has been used to produce probabilistic forecasts of severe convective storms in the United States (e.g. \citeauthor{Mecikalski_et_al_2021}, \citeyear{Mecikalski_et_al_2021}; \citeauthor{Schumacher_et_al_2021}, \citeyear{Schumacher_et_al_2021}; \citeauthor{Hill_et_al_2023}, \citeyear{Hill_et_al_2023}). \citet{Leinonen_et_al_2022} used gradient boosting for nowcasting of thunderstorm hazards in the north-eastern United States. \citet{Raedler_et_al2019} and \citet{Battaglioli_et_al_2023} developed Generalised Additive Models (GAMs) to estimate the probability of lightning and hail occurrence based on ERA-Interim (\citeauthor{Raedler_et_al2019}, \citeyear{Raedler_et_al2019}) and ERA5 (\citeauthor{Battaglioli_et_al_2023}, \citeyear{Battaglioli_et_al_2023}) predictors. The model developed in \citet{Battaglioli_et_al_2023} was also adapted for use in operational forecasting in Europe (\citeauthor{Battaglioli_et_al_2023b}, \citeyear{Battaglioli_et_al_2023b}). Likewise, \citet{Yousefnia_et_al_2024} developed a model applicable to operational forecasting using neural networks to infer the probability of lightning by feeding in predictors from high-resolution NWP data. Some studies also suggest that machine learning based-convective trigger functions could potentially replace conventional parametrizations (e.g. \citeauthor{Zhang_et_al_2021}, \citeyear{Zhang_et_al_2021}; \citeauthor{Kumar_et_al_2024}, \citeyear{Kumar_et_al_2024}). We note that this list is by no means exhaustive due to the rapid development of the field. Aside from prediction itself, these models can also be used to understand the importance of different mechanisms (features) on determining an outcome (e.g. likelihood of convection). Building trust in these models also requires a comprehensive understanding of how they arrive at their predictions, a concept sometimes referred to as explainable artificial intelligence (XAI). With the rapid evolution of field, it is crucial to evaluate the strengths and limitations of data-driven models across various applications. \\

One particularly interesting synoptic situation in which convection often initiates is in proximity to cold fronts (e.g. \citeauthor{Ferretti2014}, \citeyear{Ferretti2014}; \citeauthor{Dahl_and_Fischer_2016}, \citeyear{Dahl_and_Fischer_2016}; \citeauthor{Kunzetal2020}, \citeyear{Kunzetal2020}). This application is particularly interesting as it is known that the spatiotemporal frequency of cells in proximity to cold fronts varies substantially (\citeauthor{Pacey_etal_2023}, \citeyear{Pacey_etal_2023}). More specifically, cells are most frequent on the warm-side of the cold front (pre-frontal) and have a weakened diurnal cycle. In contrast, cells are much less frequent on the cold-side of the front (post-frontal) and have a strong diurnal cycle. The minimum cell frequency is closely correlated with the location of the 700 hPa frontal line. The skill of data-driven model predictions, similar to NWP, is likely to depend on the specific synoptic situation. For forecasters to be able to benefit from such predictions, they should be aware of the skill variation in different synoptic situations. 

To this end, an extreme gradient-boosted model (XGBoost) is trained to predict the probability of cold-frontal convective cell occurrence in Germany. Gradient boosting (\citeauthor{Friedman_2001}, \citeyear{Friedman_2001}) has already been extensively applied to several atmospheric science applications ranging from nowcasting to climate mitigation (e.g. \citeauthor{Sprenger_et_al_2017}, \citeyear{Sprenger_et_al_2017}; \citeauthor{Hsu2022}, \citeyear{Hsu2022} and \citeauthor{Leinonen_et_al_2022}, \citeyear{Leinonen_et_al_2022}).  Here, assessing the variation of skill of the data-driven model at different locations relative to the cold front is a key focus of this study. In line with previous studies, predictors are primarily obtained from data involving the use of conventional numerical models (ERA5 reanalysis data in this case). However, additional predictors that are not dependent on numerical model data, such as the time of day and latitude, are also explored. To build trust in the model outputs, feature importance techniques are also utilised.\\

The paper is organised as follows: section \ref{sec:data_and_model} introduces the data and machine learning model used in this study. Section \ref{sec:features_optimisations} presents a feature elimination strategy to automatically select features and optimise model performance. Feature importance scores also highlight the best and worst performing predictors. In section \ref{sec:model_validation}, the optimised model is validated against the observational cell climatology and potential reasons for biases are discussed with the aid of further feature importance techniques. Finally, the key findings and potential applications of the model and approach are discussed in section \ref{sec:conclusions}.

\section{Data and model}\label{sec:data_and_model}

An observational-based radar convective cell detection and tracking dataset (section \ref{subsec:KONRAD}) and ERA5 reanalysis data (\citeauthor{Hersbach2020}, \citeyear{Hersbach2020}) are combined during the summer season for 2007–2016. The period 2007–2016 is motivated by the availability of the convective cell dataset. The data are split into training and testing datasets by year (see section \ref{subsec:splitting}). ERA5 data is used for the majority of the predictors and the convective cell dataset is used as the target variable, i.e. convective cell occurrence. The model is only trained on days where cold fronts were detected in Europe. An overview of how the cold-frontal dataset is developed is given in section \ref{subsec:cold_fronts}.

\subsection{Training region}\label{subsec:training_region}

\begin{figure}
    \centering
    \includegraphics[trim={11.5cm 0cm 0cm 0cm},clip,scale=0.2]{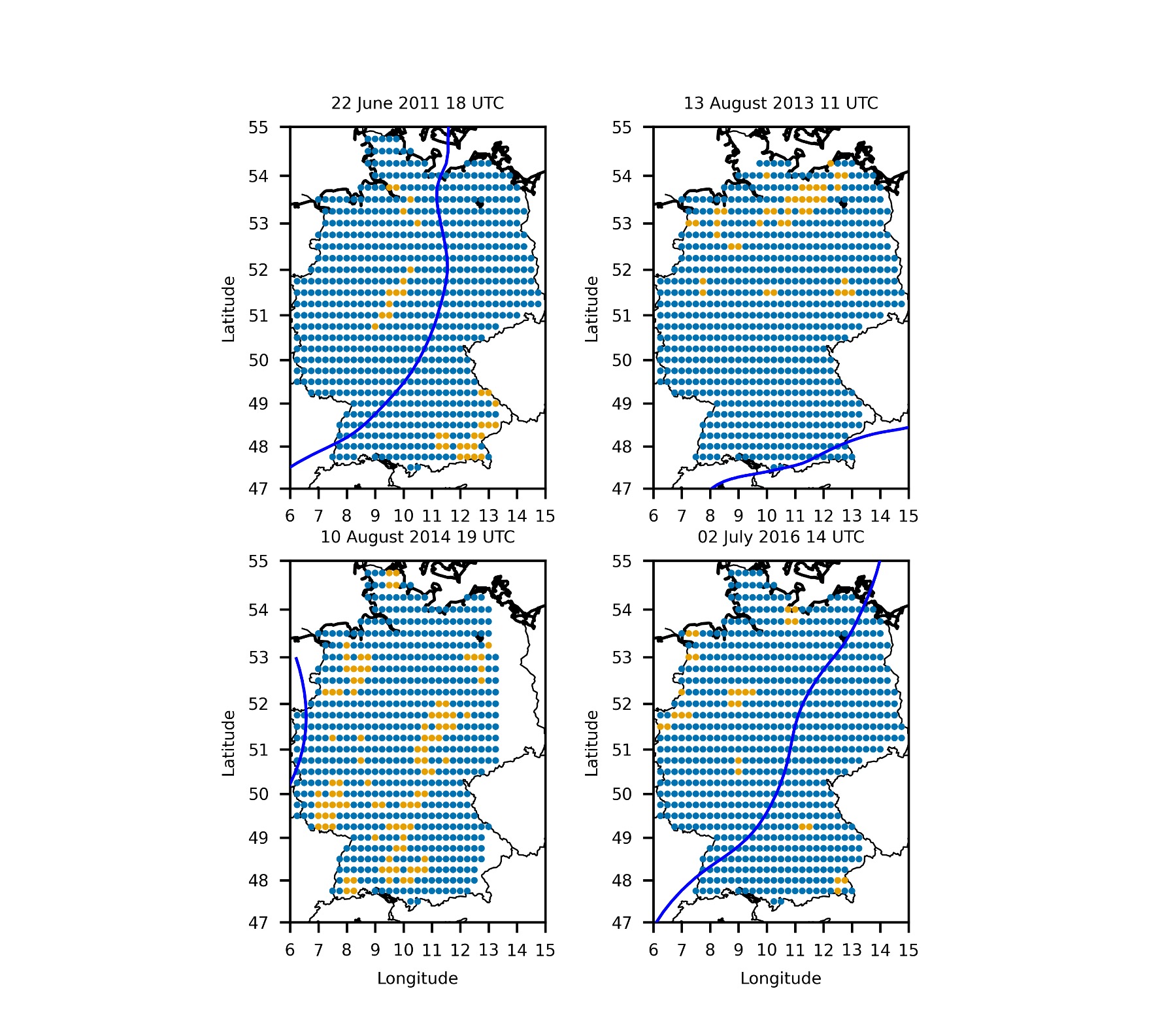}
    \caption{Four example timesteps with ERA5 grid points labelled the positive (orange) or negative class (blue). The cold frontal line at 700 hPa is shown by the blue line. Note that if ERA5 grid points are not within 500 km of the 700 hPa frontal line they are not included in the analysis (e.g. bottom-left panel)}
    \label{fig:training_region}
\end{figure}

The model training region is Germany using ERA5 grid points (Figure \ref{fig:training_region}). The cell tracking dataset does allow for cell detection outside of the borders of Germany, including into the North Sea and Baltic Sea as well as neighbouring countries. Nevertheless, the dynamics of convective cell development over the sea may differ compared to over land. Since the dataset is derived using radars in Germany only, regions outside of Germany are also towards the edge of the radar domain. Therefore, subsetting Germany reduces the number of potential missed cell detections and the number of potentially erroneously negative labelled points when there was indeed a convective cell.

\subsection{Cold-frontal dataset}\label{subsec:cold_fronts}

The cold-frontal dataset produced by \citet{Pacey_etal_2023} is used to locate whether ERA5 grid points are in proximity to cold fronts. The cold-frontal dataset is produced by applying automatic front detection methods (\citeauthor{Hewson1998}, \citeyear{Hewson1998}) to ERA5 reanalysis data. ERA5 grid points are considered associated to a cold front if they are within 500 km of cold-frontal lines defined at 700 hPa. The distance between each ERA5 grid point and the 700 hPa cold-frontal line is defined as the distance between the ERA5 grid point and the nearest grid point on the 700 hPa frontal line contour (i.e. the shortest distance between the frontal line and the ERA5 grid point on the surface of an ellipsoidal model of the earth; \citeauthor{karney_2013}, \citeyear{karney_2013}). The terminologies post-700-frontal, near-700-frontal and pre-700-frontal are used to indicate regions relative to the 700 hPa front (see Table \ref{tab:definitions}). For further details on the cold front detection algorithm the reader is referred to section 2 of \citet{Pacey_etal_2023}. \\

\subsection{KONRAD Convective Cell Detection and Tracking Algorithm}\label{subsec:KONRAD}

KONRAD (KONvektionsentwicklung in RADarprodukten, convection evolution in radar products) is a convective cell detection and tracking algorithm originally applied to 2D observational radar data in the German radar domain (\citeauthor{WaplerandJames2015}, \citeyear{WaplerandJames2015}). KONRAD is run operationally by the German Weather Service (DWD) with a spatial and time resolution of 1 km and 5 minutes respectively. A convective cell is defined as an area with 15 pixels or more exceeding 46 dBZ. As the spatial resolution of KONRAD is 1 km, 1 pixel {\raisebox{0.5ex}{\texttildelow}} 1 km\textsuperscript{2}. The reflectivity is based on a 0.5 degree radar scan; thus, the height relative to the ground varies with distance from the radar and due to orography. Where two radar scans overlap, the highest dBZ value is used. The cell centre as well as the maximum north, south, west and east extent of the cell are provided. Further details are available in \citet{Pacey_etal_2023} (their section 2.2). At the beginning of this study, the newer version of KONRAD (KONRAD 3D; \citeauthor{Werner2017}, \citeyear{Werner2017}) was not available over a longer time series of data. Therefore, KONRAD3D was not suitable for training a machine learning model. Here, the KONRAD2D dataset is used where the the convective cell definition is solely based on the criteria below:

\begin{center}
    1. \(\text{Reflectivity} \geq 46 \text{ dBZ}\)\\
    2. \(\text{Cell Area} \geq 15 \text{ km}^2\)
\end{center}

\subsection{Labelling}\label{subsec:labelling}

In this study, a probabilistic model is developed which requires labelling of positive and negative classes. ERA5 grid points at each timestep are labelled either the positive or negative class depending on whether a convective cell is associated to that grid point. Temporally, cells are associated to the timestep before the first cell detection time. For example, if a cell is first detected between 16:00–16:59 UTC, the cell would be assigned to the 16 UTC ERA5 timestep. Applying this approach avoids sampling the post-convective environment where the instability has already been released. Spatially, ERA5 grid points within the maximum north, south, west and eastern extent of the cell area are labelled as convective cell grid points. Since some cells have a lower area than the grid size the bounds are increased by 0.125 degrees (half a grid point) to ensure every cell is associated to at least one grid point. In other words, grid points falling within the cell bounds extended by 0.125 degree are labelled as the positive class. Applying this approach ensures that the general area where convection occurs is assigned the positive class. A visualisation of the class labelling is shown for a case study in August 2014 in Figure \ref{fig:labelling}. This case study is also present in Figure \ref{fig:training_region}.

\subsection{Splitting}\label{subsec:splitting}

Using ERA5 grid points within 500 km of the 700 hPa cold front in Germany during June, July and August of 2007–2016 yields 2,738,089 ERA5 grid points available for model development. While a 70/30 or 80/20 training/test split is common, if 7 years were selected as training this would only leave 3 years of testing data. Since the primary aim is to evaluate the model’s ability to reproduce the climatology, 5 years are used for training and testing. Odd years are used for training: 2007, 2009, 2011, 2013, 2015 and even years are used for testing: 2008, 2010, 2012, 2014 and 2016, to ensure independent datasets. To select the optimal number of features and hyperparameters, the training dataset is further divided into different validation folds. 
The datasets are summarised below in Table \ref{tab:data_train_test}. We note that the dataset is highly imbalanced, in other words, there are many more negative events than positive events.  Class imbalance can be an issue in machine learning since a model could end up being biased towards the majority class (e.g. \citeauthor{Krawczyk2016}, \citeyear{Krawczyk2016}). For this reason, class imbalance tests are performed to assess if the model performance is significantly affected by this imbalance (see section \ref{subsec:hyperparameter}).


\subsection{Extreme Gradient Boosting}\label{sectio}

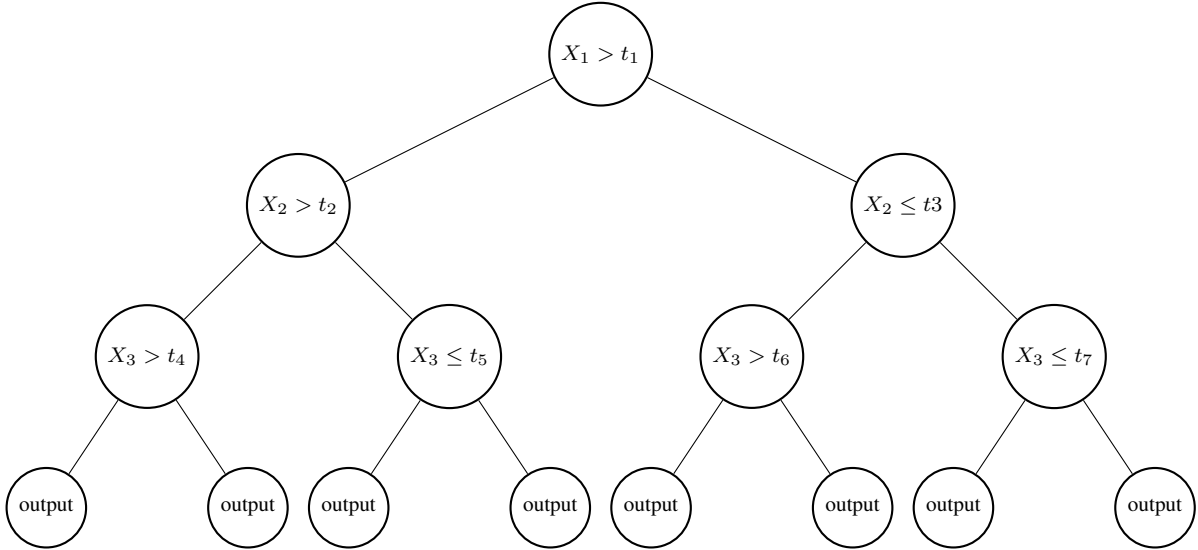
\begin{figure}
\centering
\begin{tikzpicture}[
  grow=down,
  every node/.style={draw, circle, fill=white, thick, minimum size=6mm,font=\fontsize{8}{12}\selectfont},
  level/.style={sibling distance=80mm/#1},
  level distance=20mm
]
\node {\(X_1 > t_1\)}
  child { node {\(X_2 > t_2\)}
    child { node {\(X_3 > t_4\)}
      child { node {output} }
      child { node {output} }
    }
    child { node {\(X_3 \leq t_5\)}
      child { node {output} }
      child { node {output} }
    }
  }
  child { node {\(X_2 \leq t3\)}
    child { node {\(X_3 > t_6\)}
      child { node {output} }
      child { node {output} }
    }
    child { node {\(X_3 \leq t_7\)}
      child { node {output} }
      child { node {output} }
    }
  };
\end{tikzpicture}
\caption{An example decision tree of depth 3 with conditions based on features \(X_1\), \(X_2\), and \(X_3\), where \(t_n\) are threshold values. The final prediction is obtained by summing the output scores (leaves) from all trees and then applying a link function to obtain a probability between 0 and 1.}
\label{fig:decision_tree_xgboost}
\end{figure}

Extreme Gradient Boosting (often simply referred to as XGBoost) is an optimised implementation of gradient boosting (\citeauthor{Friedman_2001}, \citeyear{Friedman_2001}), a machine learning technique which builds a predictive model by combining the predictions of multiple decision trees (see example in Figure \ref{fig:decision_tree_xgboost}). Each decision tree can be built with a different number of predictors rather than using all predictors. XGBoost iteratively improves the model's overall performance by attempting to correct the errors made by the previous trees. The number of iterations and number of trees constructed is usually referred to as boosting rounds. The predictive power of gradient boosting lies not in any individual tree but rather the ensemble of all trees. Here, the loss is minimised using binary cross entropy. The model is selected primarily due to its trade off between skill and computational expense. XGBoost is very computationally efficient compared to other tree-based approaches while often exhibiting comparable skill (\citeauthor{Bentejac_et_al_2019}, \citeyear{Bentejac_et_al_2019}). Decision-tree approaches such as Random Forest and XGBoost have been sometimes shown to outperform deep learning approaches on tabular data, especially on datasets with around 10,000 samples (\citeauthor{Grinsztajn_et_al_2022}, \citeyear{Grinsztajn_et_al_2022}). An advantage of the model presented here is that it can be easily trained and tested on a single central processing unit (CPU). Nonetheless, multiple CPUs in a high-performance computing (HPC) cluster were still used for the hyperparameter tuning. 

\section{Features and model optimisations}\label{sec:features_optimisations}

This section introduces the feature selection process and feature importance. Following the feature selection, a hyperparameter tuning phase is performed as well as class imbalance tests (see section \ref{subsec:hyperparameter}). 

\subsection{Feature selection procedure}

It is well-established that deep moist convection in the atmosphere can only occur if three ingredients are available: moisture, lift and instability (\citeauthor{Doswell1996}, \citeyear{Doswell1996}). It is assumed that the majority of the convective cells detected in KONRAD originate from deep moist precipitating clouds. On that premise, a total of 60 predictors are initially selected (Table \ref{table:full_list_of_predictors}). These include traditional predictors used in studies on the environments of convective storms, e.g. convective available potential energy (CAPE), lapse rates, vertical wind shear and relative humidity (\citeauthor{Taszarek2020Part2}, \citeyear{Taszarek2020Part2}). Several additional predictors are included which relate specifically to the lifting (triggering) mechanism. For example, Q-vector convergence, vertical velocity and total solar incoming radiation in ERA5 are also tested. Furthermore, predictors not from reanalysis data are also tested. The elevation above mean sea level is used as a proxy for topographic influences on convection (\citeauthor{Kirschbaum_et_al_2018}, \citeyear{Kirschbaum_et_al_2018}). Since convection may occur at lower elevations next to higher elevations (e.g. downslope) the distance from the nearest elevation exceeding 200, 500 and 800 metres are considered (ELEV200, ELEV500, ELEV800). Since convection is known to exhibit a diurnal cycle, the time of day is also included as a cosine transformation (cos$_{time}$; Figure \ref{fig:cosine_time_function}). The $u$ and $v$ components of wind and wind speed at different pressure levels are tested since wind direction and speed influence moisture advection and wind shear. The full list of predictors and their abbreviations are shown in the appendix (Table \ref{table:full_list_of_predictors}). 

\begin{figure}
    \centering
    \includegraphics[trim={0cm 0cm 0cm 0cm},clip,scale=0.5]{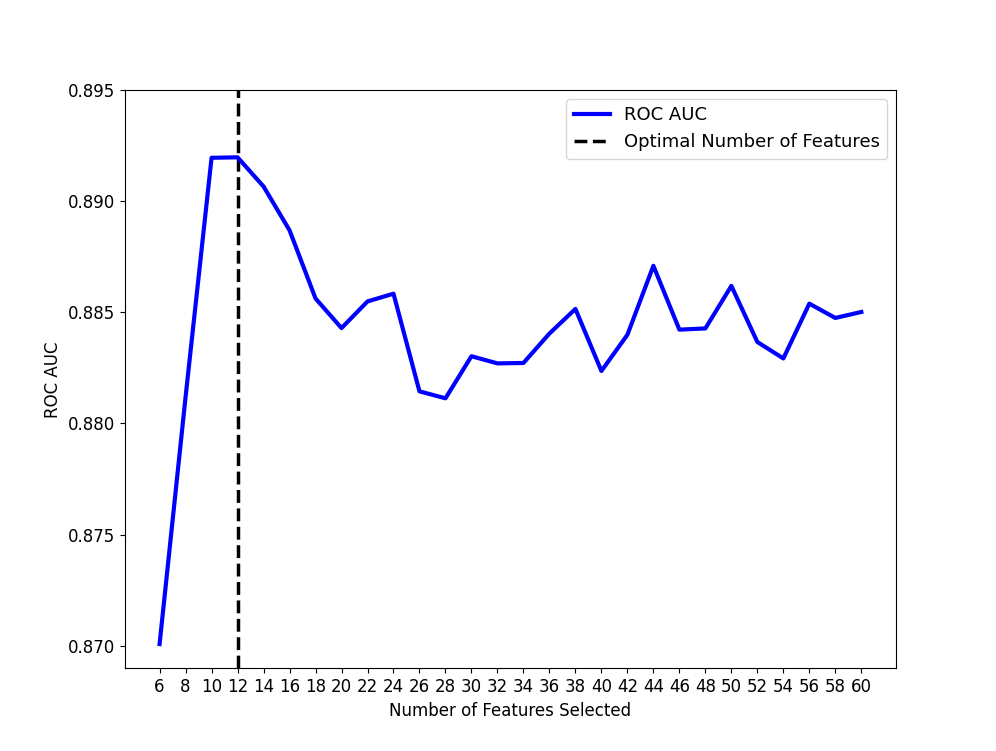}
    \caption{ROC AUC depending on the number of features selected during the RFECV.}
    \label{fig:RFECV_num_features}
\end{figure}

Training XGBoost with all 60 features may add noise to the model if redundant features are included. Furthermore, using a larger number of features comes with increasing computational cost. The optimal number of features is selected using Recursive Feature Extraction with Cross-Validation (RFECV; \citeauthor{Kuhn_and_Johnson_2013}, \citeyear{Kuhn_and_Johnson_2013}). The model is initially trained with all features (60 in this case), features are ranked from most to least important, and the least important features are then removed. These steps are repeated until the minimum number of features is reached. A step of 2 is used meaning 2 features are removed during each round. RFECV selects the number of features that maximise the model's performance based on the cross-validation result, that is to say, using different training and testing datasets. The model's performance and optimal number of features is selected based on the receiver operating characteristic area under the curve (ROC AUC). The minimum number of features is predefined as 6 due to previous domain knowledge that multiple factors are relevant for the development of convective cells. Indeed, the model skill reduces steeply after 10 or less features (Figure \ref{fig:RFECV_num_features}). The best 12 features and their feature importance are shown in Table \ref{tab:thermodynamic_kinematic_other} and Figure \ref{fig:best12_feature_importance}. The feature importance is derived using the built-in functionality of XGBoost which is based on how much each feature contributes towards reducing the model error during the splitting of tree nodes (Gini impurity; \citeauthor{breiman_et_al_1984}, \citeyear{breiman_et_al_1984}). The ROC AUC depending on the number of features is shown in Figure \ref{fig:RFECV_num_features}. The ranking for all 60 features is shown in the Appendix (Figure \ref{fig:RFECV_ranking}). The RFECV process was also carried out selecting the optimal number of features by the highest precision-recall area under the curve (PR AUC). Based on the PR AUC, 10 would be the optimal number of features. Given this similarity, the selection does not significantly affect the subsequent results. \\






\subsection{Feature importance}\label{subsec:feature_importance}

CAPE is the most important predictor followed by total column water vapour and vertical velocity at 700 hPa. A study on modelling lightning occurrence in Europe using machine learning also showed instability features as most important (\citeauthor{Ukkonen_and_Makela_2019}, \citeyear{Ukkonen_and_Makela_2019}). The best 12 predictors (Table \ref{tab:thermodynamic_kinematic_other}) include several predictors related to the three ingredients for deep moist convection. The instability is represented by CAPE and also by lapse rates between 700 hPa and 500 hPa (LR75). Steeper lapse rates between 700 to 500 hPa generally allow stronger buoyancy in the mid-levels. However, LR75 is amongst the least important out of the 12 predictors (Figure \ref{fig:best12_feature_importance}), possibly owing to its partial collinearity with CAPE. 
Total column water vapour (TCWV) is a measure of the integrated water vapour content of the atmosphere. While specifically lower level moisture is generally thought to be most relevant for convective initiation, the majority of moisture is concentrated in the lower levels of the atmosphere. Correlation coefficients between TCWV and surface dew point (Td$_{2m}$), specific humidity at 900 hPa and 850 hPa are 0.79, 0.77 and 0.79, respectively. The correlation between TCWV and the mean of QHUM900, QHUM850, QHUM700 is 0.89. This high correlation indicates that the TCWV feature contains information about the low-level moisture availability. Thus, using a specific level such as QHUM900 or surface dewpoints may be less effective compared to a combined parameter across a column as air parcels can depart from different levels depending on the situation. Furthermore, it is of interest to compress information into a lower number of features since increasing the number of features adds increasing complexity (as shown in Figure \ref{fig:RFECV_ranking}). Mid-level relative humidity (RH700 and RH850–500) is also in the best 12 variable combination. Mid-level relative humidity is relevant due to the possibility of entrainment (e.g. \citeauthor{Morrison_et_al_2022}, \citeyear{Morrison_et_al_2022}). Drier environmental air may entrain into the updraught and weaken buoyancy, potentially hindering the development of convective cells.
Vertical velocity at all three levels are present in the best 12 variable combination, while Q-vector convergence does not appear. A significant difference between the mean of vertical velocity at cell grid points and locations without cells was observed in \citet{Pacey_et_al_2025} (their Figure 8). It is possible that when convection is triggered in the parameterization scheme of ERA5 that there is some feedback on the vertical velocity field. Therefore, it cannot be assumed that the importance of vertical velocity in the model indicates importance of the background synoptic- or mesoscale scale lifting, although it may play some role. Q-vector convergence does not separate as well as the vertical velocity between cell grid points and locations without cells (\citeauthor{Pacey_et_al_2025}, \citeyear{Pacey_et_al_2025}; their Figures S4 and S5). The Q-vector convergence also only contains information about the quasi-geostrophic forcing for ascent (large-scale lifting) and does not consider mesoscale lifting for which a diagnostic tool is not currently available. 
Higher surface temperatures (TEMP$_{2m}$) reduce CIN and increase CAPE (\citeauthor{trapp_2013}, \citeyear{trapp_2013}). If the higher temperatures are a result of stronger solar heating this also favours lifting of air parcels near the surface. High cloud cover is also found to be amongst the best predictors. The presence of high cirrus clouds may indicate existing deep moist convection and enhanced high cloud cover. Convection often initiates in the vicinity of pre-existing convection due to gust fronts (e.g. \citeauthor{Lima_and_Wilson_2008}, \citeyear{Lima_and_Wilson_2008}; \citeauthor{Hirt_et_al_2020}, \citeyear{Hirt_et_al_2020}). Finally, the cosine transformation of the time day (cos$_{time}$) is amongst the best predictors. It is known that convection usually exhibits a diurnal cycle, primarily due to variations in solar heating through the day. In the case of cold-frontal convective cells, there are differences in the diurnal cycle depending on the location relative to the front (\citeauthor{Pacey_etal_2023}, \citeyear{Pacey_etal_2023}; Figures 6 and 7). We will revisit the importance of this particular feature at the end of section \ref{sec:model_validation}. 


An important caveat to note is that the absence of a feature in the 12 feature model does not indicate unimportance of that feature for convective cell development. For example, some wind shear is generally present across cold fronts so may carry less importance in the model developed in this study than if the model were trained on all cases including non-cold-frontal cells. An additional note to consider is that features correlated with strong predictors may appear as less important. This is highlighted in the feature ranking for the RFECV shown in Figure \ref{fig:RFECV_ranking}. Amongst the worst predictors is w$_{max}$, however, the best performing predictor in the best 12 predictor model is CAPE (Figure \ref{fig:best12_feature_importance}). CAPE and w$_{max}$ have a Pearson correlation of 0.92. Therefore, the predictor w$_{max}$ does not add any unique information since it is just a transformation of CAPE. If CAPE were switched for w$_{max}$ in the 12 predictor model, w$_{max}$ would be the most important predictor (tested and verified). Likewise, the elevation features are removed first during the RFECV (Figure \ref{fig:RFECV_ranking}) indicating low importance. However, the elevation is negatively correlated with latitude, since higher elevation is generally in the south of Germany at lower latitudes. The reason latitude may be more important than elevation could be due to the fundamental nature of decision tree algorithms. Latitude has only 30 unique values, ranging from 47.5 to 54.75 at 0.25° intervals (matching the ERA5 spatial resolution), which may make it easier for the tree to find effective splits. In contrast, elevation is continuous with around 700 unique values, potentially making it harder for the model to identify an optimal splitting point. \\
While highlighting the most optimal features for the model, the RFECV results also emphasise the need to be cautious when drawing physical conclusions. Following the RFECV, the hyperparameters are tuned based on the best 12 features (see section \ref{subsec:hyperparameter} for further details). The final model only requires around 10 GB of RAM for both training and inference and can be performed on a local machine (assuming sufficient RAM availability).

\section{Model validation}\label{sec:model_validation}

In this section, model performance is evaluated on the yearly scale (Figure \ref{fig:yearly_and_monthly_model_obs}a), monthly scale (Figure \ref{fig:yearly_and_monthly_model_obs}b) and for different regions relative to the cold front (Figs.  \ref{fig:cell_front_dist_climo_test_data}– \ref{fig:diurnal_cycle_pre_post}). Because the number of grid points varies by year and month due to differences in cold front frequency, the results are expressed as the fraction of grid points with cells in both the model and observations. In other words, the relative frequency probability is shown. This enables a fair assessment of how well the model performs on the testing data. 

\subsection{Yearly and monthly cell count}
Overall, the relative cell frequency is modelled well with the highest and lowest years correctly identified as 2014 and 2010, respectively (Figure \ref{fig:yearly_and_monthly_model_obs}a). The model only exhibits a slight overestimation in 2008 and underestimation in 2014 and 2016.  
The overestimation in 2008 is linked to July and August (Figure \ref{fig:yearly_and_monthly_model_obs}b). In 2016, the underestimation is primarily linked to May. As on the yearly scale, the model can identify months with a high/low relative frequency of cold-frontal convective cells. The mean absolute error (MAE) for the monthly relative frequency is 0.19. As a baseline, a single predictor logistic regression model is trained and tested with ERA5 convective precipitation as the only predictor. Convective precipitation above zero is present in ERA5 when the physics-based parametrization scheme is triggered. ERA5 is based on the Intergrated Forecast System (IFS) Cycle 41r2, thus the same convective trigger function is used as in IFS. The convective precipitation accumulated during the hour after the ERA5 timestep is used to match the period used for labelling grid points as convective cells (see section \ref{subsec:labelling}). A weak agreement between modelled and observed cell frequency was found on the monthly scale with a MAE of 0.43. The optimised 12 predictor model outperforms the baseline model across a range of metrics (see Table \ref{table:skill_scores}). 

\subsection{Model skill relative to the cold front}

The model skill is now assessed depending on the region relative to the cold front. In observations, the maximum relative cell frequency is in the region 250–300 km ahead of the 700 hPa front and lowest frequency surrounding the 700 hPa front (Figure \ref{fig:cell_front_dist_climo_test_data}). \citet{Pacey_etal_2023} argue that the maximum relative cell frequency is linked to the surface front, which is located around 300 km ahead of the 700 hPa front on average based on mean cold frontal slopes (1:100; \citeauthor{Bott2016_slope}, \citeyear{Bott2016_slope}). This assumption is also supported by the mean surface convergence in ERA5 (\citeauthor{Pacey_etal_2023}, \citeyear{Pacey_etal_2023}; Figure 3). The model also shows the highest relative cell frequency in the region 250–300 km but this is underestimated compared to observations. This result serves as a guidance for forecasters using ML-based forecasts of convection since the occurence near the surface front may be somewhat underestimated. The underestimation could be linked to convective organisation near the surface front where lifting due to outflow boundaries and cell interactions (e.g. \citeauthor{Hirt_et_al_2020}, \citeyear{Hirt_et_al_2020}) are not explicitly considered in the predictors. There is also an overestimation of the relative cell frequency in the region 0–100 km. Behind the 700 hPa front (post-700hPa-frontal), a slight increase in the relative cell frequency is observed, which is also shown by the model. \\

Numerical models continue to exhibit biases in representing the diurnal cycle of convection (e.g., \citeauthor{Tao_et_al_2024}, \citeyear{Tao_et_al_2024}), although data-driven approaches have the potential to mitigate these biases. The diurnal cycle of convective cells is shown in Figure \ref{fig:diurnal_heatmap_climo_test_data} for the observations (left) and model (right) depending on the distance from the front. On the post-700hPa-frontal side, the model correctly represents a low relative cell frequency during the night-time with cells mostly being confined to the daytime. The model also captures the time of day when pre-700hPa-frontal cells occur most frequently (between 16–20 UTC). A weakened diurnal cycle is also evident on the warm-side of the 700 hPa front. Nonetheless, the underestimation in the relative cell frequency is apparent near the surface front, as observed in Figure \ref{fig:cell_front_dist_climo_test_data}. Even though there are distinct temporal cell frequencies depending on the region relative to the front, the model is able to represent these differences and thus exhibits good generalisation ability. 


\subsection{Influence of temporal features on diurnal cycle bias}

We revisit feature importance to highlight how sensitive the representation of the diurnal cycle is to the time of the day feature (cos$_{time}$) in the model. To better highlight the diurnal cycle, we only show the post-700hPa-frontal (-500 to 0 km) and pre-700hPa-frontal (0 to 500 km) regions, opposed to presenting the data in 50 km bins (as done in Figure \ref{fig:diurnal_heatmap_climo_test_data}). In observations (blue line on Figure \ref{fig:diurnal_cycle_pre_post}), a strong diurnal cycle on the post-700hPa-frontal side of the front is evident (Figure \ref{fig:diurnal_cycle_pre_post}a). This is represented relatively well by the 12 predictor model (red line with circle marker) which includes cos$_{time}$. However, when the model is trained and tested without the cos$_{time}$ feature (solid red line), the cell frequency peaks too early and has a flatter peak. Convection peaking too early is a common bias in numerical models (e.g. \citeauthor{Bechtold_et_al_2014}, \citeyear{Bechtold_et_al_2014}; their Figure 5). A negative bias in the relative cell frequency is evident between 12–20 UTC. On the pre-700-frontal side, observations show a weakened diurnal cycle (Figure \ref{fig:diurnal_cycle_pre_post}b). The representation of this diurnal cycle is also inferior when the cos$_{time}$ feature is omitted. Specifically, the 11 predictor model exhibits a larger positive bias of the relative cell frequency during the morning hours but a larger negative bias during the cell maximum (12–20 UTC) compared to the 12 predictor model which includes cos$_{time}$. For comparison, we also trained and tested the model without CAPE (Figure \ref{fig:diurnal_cycle_pre_post_without_CAPE}). The diurnal cycle does not largely deviate from the observations when CAPE is omitted. The cosine transformed time of day feature seems to provide a bias correction on both sides of the cold front where there are distinct diurnal cycles. The model does not simply correct for a too early diurnal peak. Therefore, we argue that the model may have also learned the typical convective environments on both sides of the cold front (\citeauthor{Pacey_et_al_2025}, \citeyear{Pacey_et_al_2025}). Overall, the bias correction appears to be stronger on the cold side of the front. We therefore repeated the feature importance analysis in \ref{subsec:feature_importance} by training one model with pre-700hPa-frontal data (warm side of front) and another model with post-700hPa-frontal data (cold side of front) data. The cos$_{time}$ feature is indeed more important to model predictions on the cold side of the front (Figure \ref{fig:two_models_feature_importance}), consistent with what we observed on Figure \ref{fig:diurnal_cycle_pre_post}. \\
Finally, to better understand the origins of these biases when the cos$_{time}$ feature is not present, we also train a model with lead and lag predictors of CAPE and vertical velocity at 500 and 700~hPa for $\pm1$, $\pm2$, $\pm3$~h relative to the reference time. The motivation for this is that the bias may be linked to biases in the timing of convective initiation in ERA5, meaning the machine learning model sees too much or not enough instability. The same applies to vertical velocity, since the parameterization scheme being triggered is likely to feedback on the vertical velocity field in ERA5. The dashed red line on Figure \ref{fig:diurnal_cycle_pre_post} shows a model trained with the 11 best predictors (only excluding cos$_{time}$) but with 18 additional lead and lag features (29 features in total). The diurnal cycle on both sides of the front is indeed better represented than the 11 predictor model but not as good as the 12 predictor model which includes cos$_{time}$. We conclude that the diurnal cycle bias with the 11 predictor model is partly due to the misrepresentation of the timing of convection in ERA5 and including the time of day as an input predictor best helps to mitigate this. We also note that the RFECV and hyperparameter tuning were not repeated after the inclusion of lead and lag features so additional performance improvements may still be achievable by incorporating these features at the start of the pipeline.
These results provide valuable guidance for future ML model development on convection-related applications, indicating that static features as well as lead and lag features may enhance model performance. 


\section{Concluding remarks}\label{sec:conclusions}


A gradient-boosted model was trained to predict the occurrence of cold-frontal convective cells in Germany during the summer season between 2007–2016. The predictors primarily came from ERA5 but additional predictors such as the elevation and the time of day were also tested. A total of 60 predictors were initially considered and a feature elimination technique as well as hyperparameter tuning were utilised to produce an optimised model for cold-frontal convective cells. The resulting model was applied to new and unseen testing data to see how well the model could reproduce the cold-frontal convective cell climatology. \\

\begin{figure}
    \centering
    \includegraphics[trim={0cm 0cm 0cm 2cm},clip,scale=0.7]{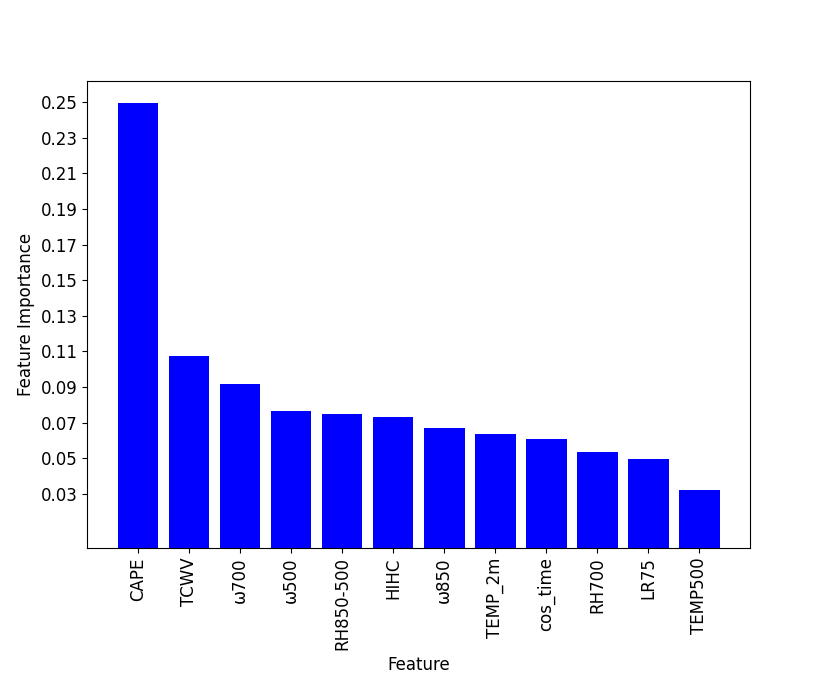}
    \caption{Summary of the best 12 features and feature importance to the model using based on Gini impurity (\citeauthor{breiman_et_al_1984}, \citeyear{breiman_et_al_1984}). Higher numbers indicate higher feature importance. Predictors are ranked from most important to least important from left to right. The feature importance values are normalised so that they sum to one.}
    \label{fig:best12_feature_importance}
\end{figure}

\begin{figure}
    \centering
    \subfigure{
        \includegraphics[trim={0cm 0cm 0cm 0cm},clip,scale=0.5]
        {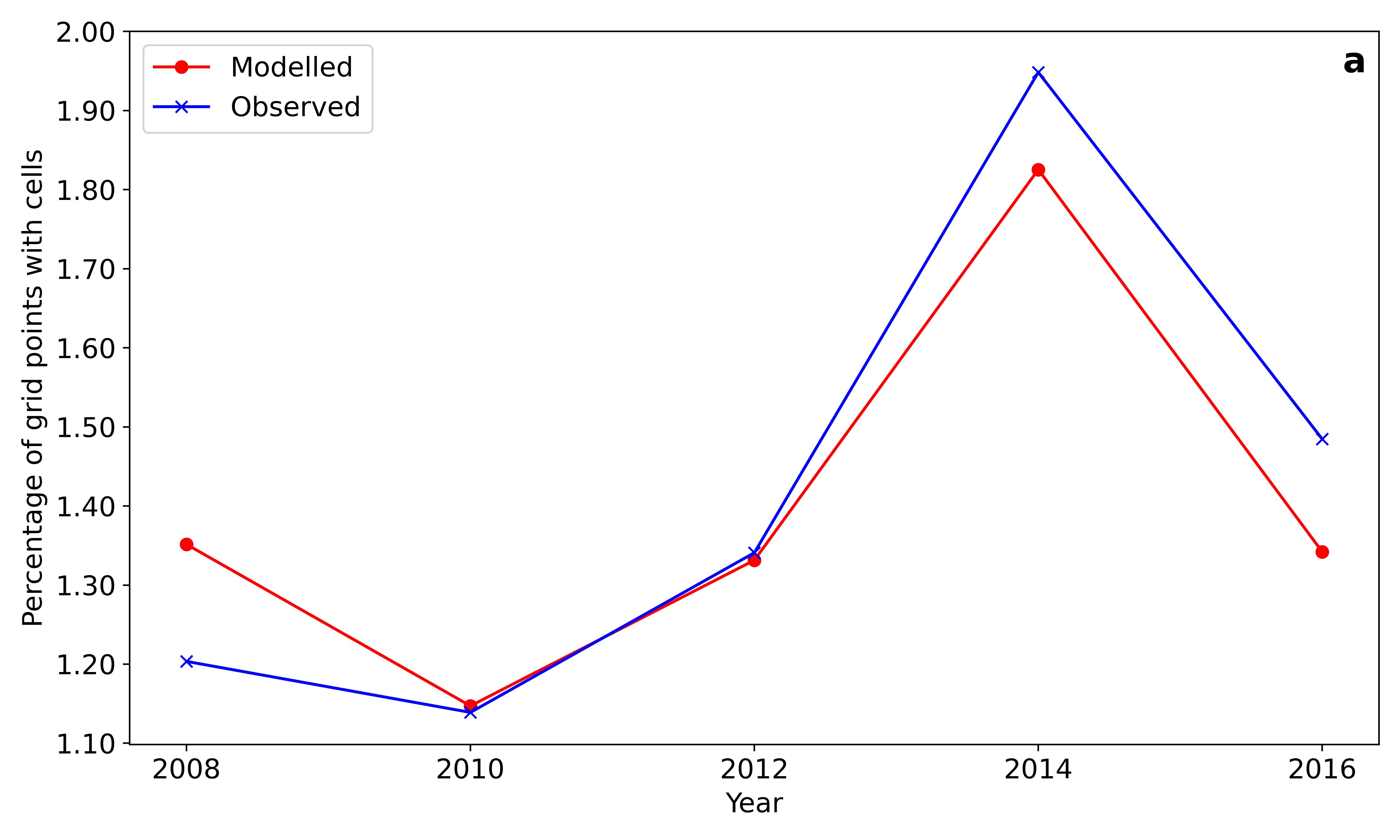}
    }
    \subfigure{
        \includegraphics[trim={0cm 0cm 0cm 0cm]},clip,scale=0.5]
        {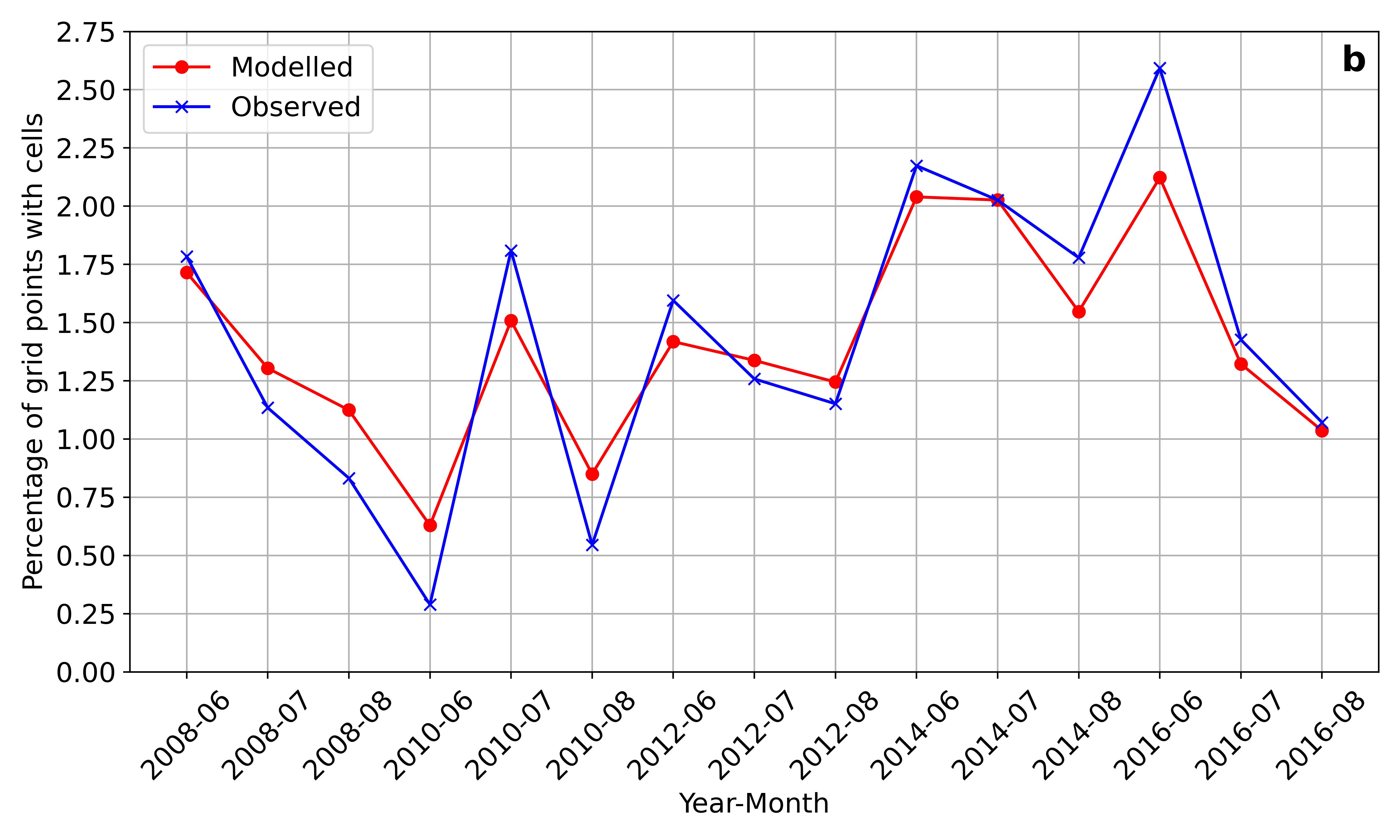}
    }
    \caption{Yearly cell count (a) and monthly cell count (b) for the model (red) and observations (blue).}
    \label{fig:yearly_and_monthly_model_obs}
\end{figure}

\begin{figure}
    \centering
    \includegraphics[trim={0cm 0cm 0cm 0cm},clip,scale=0.65]{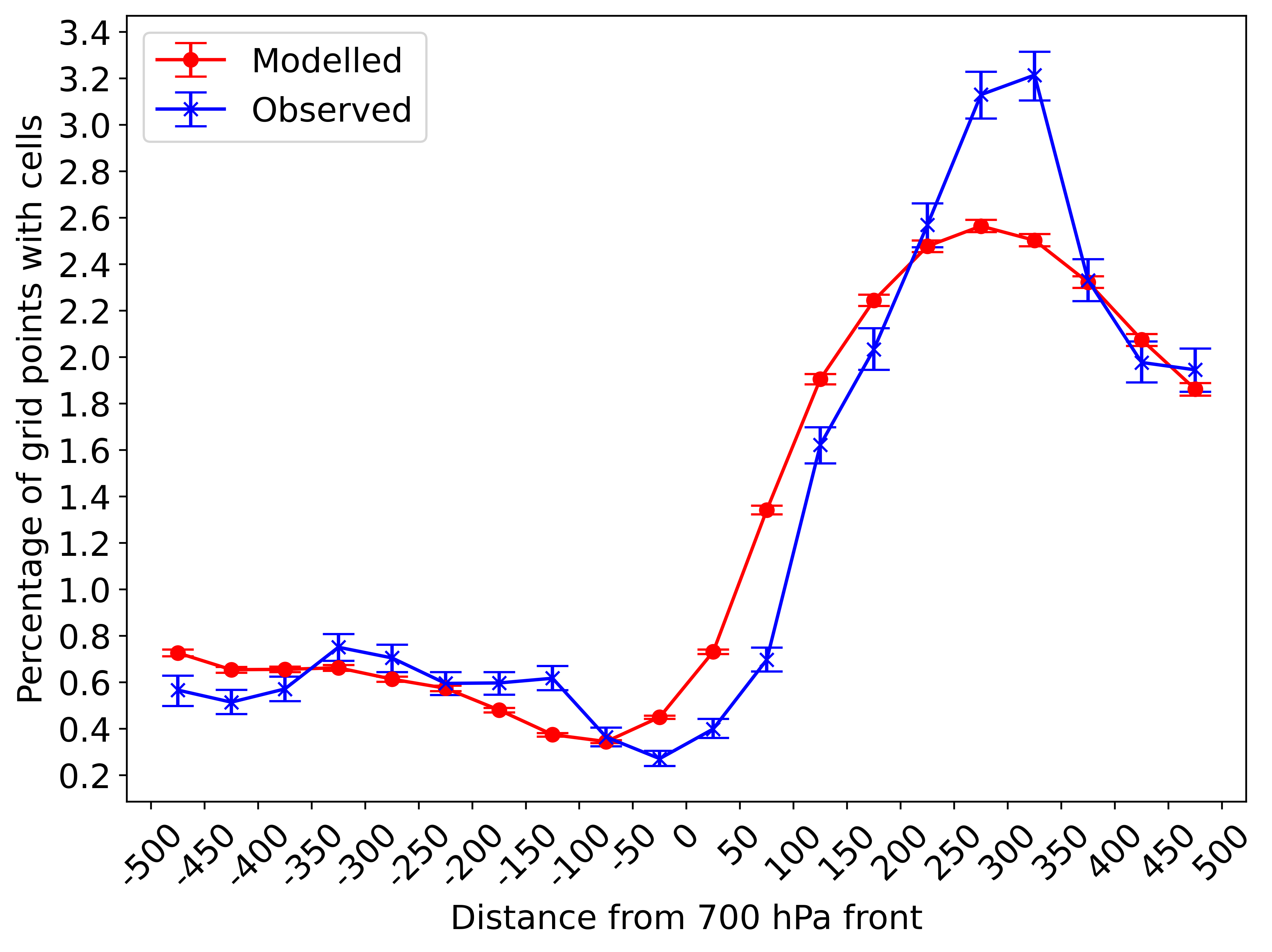}
    \caption{Cell count depending on the distance from the 700 hPa front for the model (red) and observations (blue). Error bars indicate the 5th and 95th percentiles based on 500 bootstrapped samples from the testing data. Testing years: 2008, 2010, 2012, 2014 and 2016.}
    \label{fig:cell_front_dist_climo_test_data}
\end{figure}

\begin{figure}
    \centering
    \includegraphics[trim={2.5cm 6.5cm 1cm 8cm},clip,width=\textwidth]{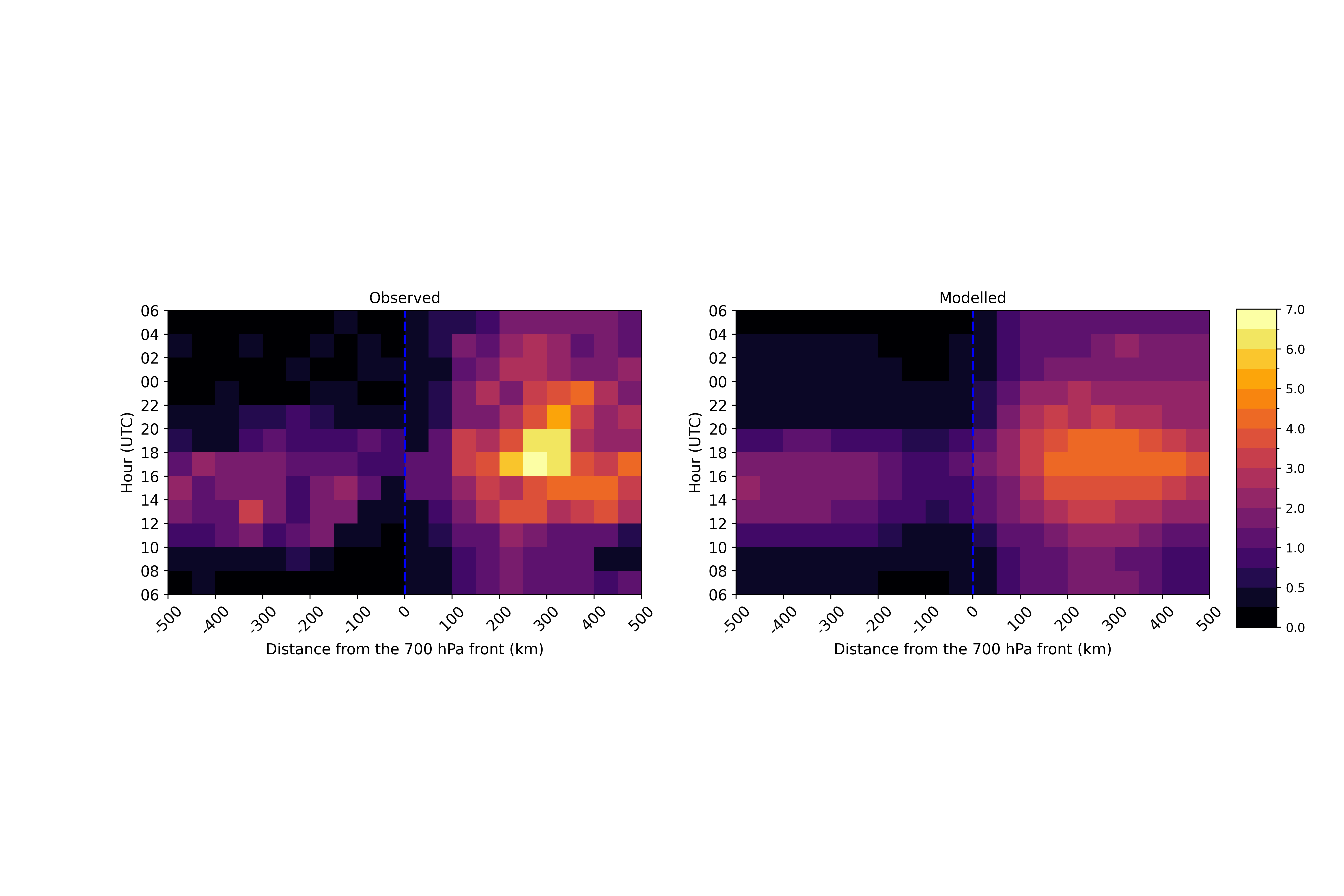}
    \caption{Cell count (non-linear colorbar) depending on the cell-front distance (horizontal axis) and hour of the day (vertical axis) for the observations (left) and model (right). The blue horizontal dashed line represents the 700 hPa front location. Testing years: 2008, 2010, 2012, 2014 and 2016.}
    \label{fig:diurnal_heatmap_climo_test_data}
\end{figure}

\begin{figure}
    \centering
    \includegraphics[trim={1cm 2cm 1cm 3cm},clip,scale=0.6]{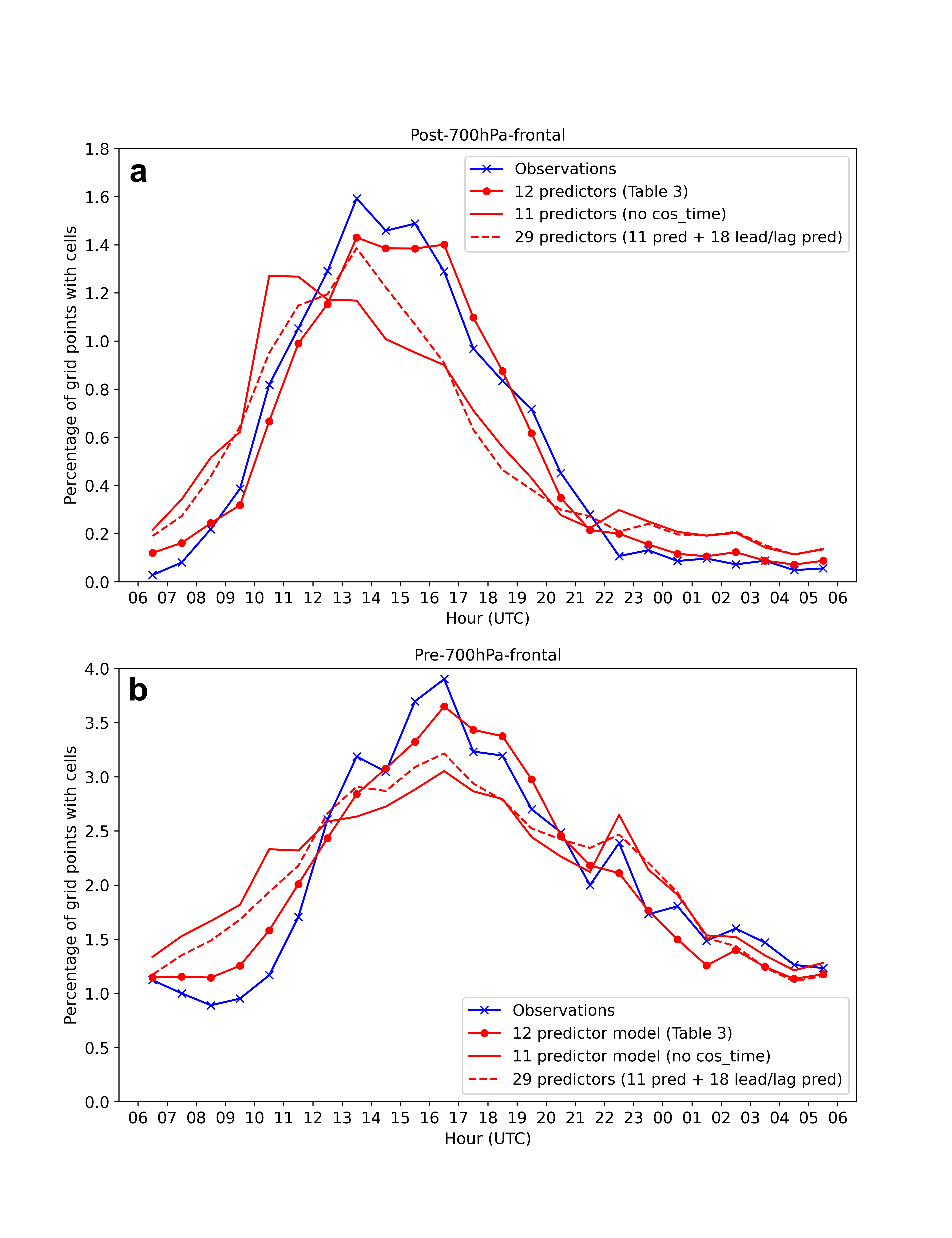}
    \caption{Diurnal cycle cell count for post-700hPa-frontal cells (a) and pre-700hPa-frontal cells (b). Observations, 12 predictor model, 11 predictor model and 29 predictor model shown by the blue, solid red line with circle markers, solid red line and dashed red line, respectively. Testing years: 2008, 2010, 2012, 2014 and 2016.}
    \label{fig:diurnal_cycle_pre_post}
\end{figure}

\noindent The primary findings from this study are highlighted below:\\
\vspace{-0.5cm}
\begin{itemize}


\item A feature elimination strategy yielded 12 as the optimal number of features (predictors), including several thermodynamic variables such as CAPE, total column water vapour and mid-level relative humidity (Table \ref{tab:thermodynamic_kinematic_other}). Vertical velocity at 3 different pressure levels, the time of day (cos$_{time}$) and high cloud cover fraction were also amongst the strongest predictors. 

\item CAPE was the most important predictor followed by total column water vapour and vertical velocity at 700 hPa (Figure \ref{fig:best12_feature_importance}), consistent with the ingredients required for deep moist convection (\citeauthor{Doswell1996}, \citeyear{Doswell1996}). 

\item The model exhibited high skill in representing the monthly and yearly cold-frontal cell frequency (Figure \ref{fig:yearly_and_monthly_model_obs}). The region where convective cells are most and least frequent relative to the front was also represented well (Figure \ref{fig:cell_front_dist_climo_test_data}). However, some biases exist in the cell frequency for certain regions relative to the front (e.g. underestimation near the surface front). Knowledge of these front-relative biases may help forecasters in judging the trustworthiness of data-driven predictions of convection during different phases of cold-frontal passages.

\item The model also represented the diurnal cycle of convective cells well on both sides of the cold front (Figure \ref{fig:diurnal_heatmap_climo_test_data}). The time of day predictor (cos$_{time}$) was shown to be particularly important for accurately modelling the diurnal cycle (Figure \ref{fig:diurnal_cycle_pre_post}). Without this predictor, the diurnal cycle peaked too early on the post-frontal side and had a flatter peak on the pre-frontal side. 

\item The optimised model outperformed a baseline model using ERA5 convective precipitation across a range of skill metrics. The finding is interesting as ERA5 convective precipitation is dependant on a conventional convective trigger function.

\end{itemize}

Despite the model not being trained to predict convective cells in a specific region relative to the front and being trained on convective cells with a wide range of environments (\citeauthor{Pacey_et_al_2025}, \citeyear{Pacey_et_al_2025}), we observe that the model does have generalisation capability. The results are very encouraging in light of using data-driven approaches to model the occurrence of convection. The model developed here has various applications within the field of convection. For example, models developed using ERA5 data as predictors can be used to reconstruct climatological trends of convective events (e.g.  \citeauthor{Battaglioli_et_al_2023}, \citeyear{Battaglioli_et_al_2023}; \citeauthor{Wilhelm_et_al_2024}, \citeyear{Wilhelm_et_al_2024}). Applying the model to post-process NWP model output (e.g. \citeauthor{Hill_et_al_2023}, \citeyear{Hill_et_al_2023}; \citeauthor{Yousefnia_et_al_2024}, \citeyear{Yousefnia_et_al_2024}) may be used to improve conventional convection forecasts. On the other hand, the probabilities could be converted to a binary class label (true or false) above a given probability threshold. In this case, such a model could be used as a convective trigger function. Some initial work has already begun in this area for climate models (e.g. \citeauthor{Zhang_et_al_2021}, \citeyear{Zhang_et_al_2021}; \citeauthor{Kumar_et_al_2024}, \citeyear{Kumar_et_al_2024}). There is also potential for machine learning-based trigger functions to be used in short- to medium-range NWP models as well. The current state-of-the-art ML weather forecast models (e.g. GraphCast, PanguWeather, AIFS) were all trained on ERA5 initially (as in this study) and then fine-tuned to be used as an operational forecast model. \\
Aside from the promising performance of the model, the feature importance and elimination strategy also builds trust in these approaches as the most important features are physically meaningful. Previous theory tells us that in order for deep moist convection to develop in space and time, three ingredients must be met: moisture, lift and instability (\citeauthor{Doswell1996}, \citeyear{Doswell1996}). The two most important features in the model are instability (CAPE) and moisture (total column water vapour). Features related to lift are also present in the model such as the vertical velocity at 3 different pressure levels. The time of day feature is correlated with insolation near the surface giving parcels energy to rise near the surface. However, we note some caveats when using feature importance techniques that should be considered. As discussed in section \ref{subsec:feature_importance}, a feature exhibiting low importance to the model does not necessarily indicate low overall importance on determining an outcome (e.g. convective cell occurrence). Correlations between other features involved in the feature importance process as well as the specific application must also be considered. 
\\Finally, we note that the general methodology followed in this study is applicable to the broader field of the atmospheric sciences. Given the computational efficiency of extreme gradient boosting, the methodology can be applied without the need for large computational resources or Graphic Processing Units. \\

\clearpage

\clearpage

\begin{table}
\caption{Definitions of regions relative to the front}
\centering
\begin{tabular}{|l|c|c|c|}
\hline
\textbf{Terminology} & \textbf{Definition} \\
\hline
Post-700-frontal & behind (on the cold-side of) the 700~\text{hPa} front \\
Near-700-frontal & within 50~\text{km} of the 700~\text{hPa} frontal line\\
Pre-700-frontal & ahead (on the warm-side) of the 700~\text{hPa} front \\
\hline
\end{tabular}
\label{tab:definitions}
\end{table}

\begin{table}
    \centering
    \caption{Summary of data used for training and testing.}
    \begin{tabular}{c|c|c|c}  
        & Training (odd years) & Testing (even years) & All (2007–2016) \\ 
        \hline
        Data points   & 1,421,902   & 1,316,187 & 2,738,089   \\ 
        Positive class (\%)   & 1.34   & 1.38 &  1.36 \\ 
        Negative class (\%) & 98.66 & 98.62 & 98.64
    \end{tabular}\label{tab:data_train_test}
\end{table}

\begin{table}
\caption{Summary of the best 12 features categorised into thermodynamic, kinematic and other. Predictor full names are shown in Table \ref{table:full_list_of_predictors}.}
\centering
\begin{tabular}{|c|c|c|}
\hline
\textbf{Thermodynamic} & \textbf{Kinematic} & \textbf{Other} \\
\hline
$T_{2\text{m}}$              & $w_{\text{850 hPa}}$ & $\cos_{\text{time}}$ \\
$T_{500 \text{ hPa}}$        & $w_{\text{700 hPa}}$ & HIHC \\
$LR_{700-500 \text{ hPa}}$   & $w_{\text{500 hPa}}$ &  \\
CAPE                         &                      &  \\
$RH_{850-500 \text{ hPa}}$   &                      &  \\
$RH_{700 \text{ hPa}}$       &                      &  \\
TCWV                         &                      &  \\
\hline
\end{tabular}
\label{tab:thermodynamic_kinematic_other}
\end{table}

\begin{table}
    \centering
    \small
    \caption{Skill metrics for the baseline logistic regression model with convective precipitation and the optimised gradient-boosted model. Note that for the ROC AUC and precision-recall (PR) AUC higher is better and for the Brier Score and MAE lower is better. A model predicting randomly would yield ROC AUC, PR AUC and Brier Score values of 0.500, 0.0138 and 0.333, respectively.}
    \begin{tabular}{c|c|c}
        & \textbf{12 predictor model} & \textbf{Baseline model with convective precip.}\\
        \hline
        \textbf{ROC AUC} & \textcolor{black}{0.899} & \textcolor{black}{0.714}
        \\
        \hline
        \textbf{PR AUC} & \textcolor{black}{0.122} & \textcolor{black}{0.0588}\\
        \hline
        \textbf{Brier Score} & \textcolor{black}{0.0128} & \textcolor{black}{0.0135}\\
        \hline         
        \textbf{Monthly relative frequency MAE} & \textcolor{black}{0.19} & \textcolor{black}{0.43}\\
    \end{tabular}\label{table:skill_scores}
\end{table}

\clearpage

\section*{Acknowledgments}

This research has been funded by Deutsche Forschungsgemeinschaft (DFG) through grant CRC 1114 “Scaling Cascades in Complex Systems, Project Number 235221301” and project C06 “Multi-scale structure of atmospheric vortices”. The authors would like to thank the HPC Service of FUB-IT, Freie Universität Berlin, for computing time (\citeauthor{Bennett_et_al_2020}, \citeyear{Bennett_et_al_2020}). We would also like to acknowledge Henning Rust and Nico Becker for useful discussions regarding verification metrics and probabilistic modelling, Akmal Khikmatullaev for useful discussions regarding gradient boosting, the DWD for providing the KONRAD convective cell tracking dataset as well as ECMWF and the Climate Data Store for the ERA5 data. 

\section*{Data availability statement}
ERA5 data can be downloaded from the Copernicus servers (\citeauthor{Hersbach2020}, \citeyear{Hersbach2020}). The KONRAD dataset is available for research purposes on request (contact kundenservice@dwd.de).

\section*{Appendix}

\appendix

\renewcommand{\thesection}{A.\arabic{section}}
\setcounter{section}{0}

\counterwithin{figure}{section}
\counterwithin{table}{section}

\renewcommand{\thefigure}{A\arabic{figure}}
\renewcommand{\thetable}{A\arabic{table}}

\setcounter{figure}{0}
\setcounter{table}{0}

\begin{figure}[H]
    \centering
    \includegraphics[trim={3cm 1cm 0cm 2cm},clip,scale=0.40]{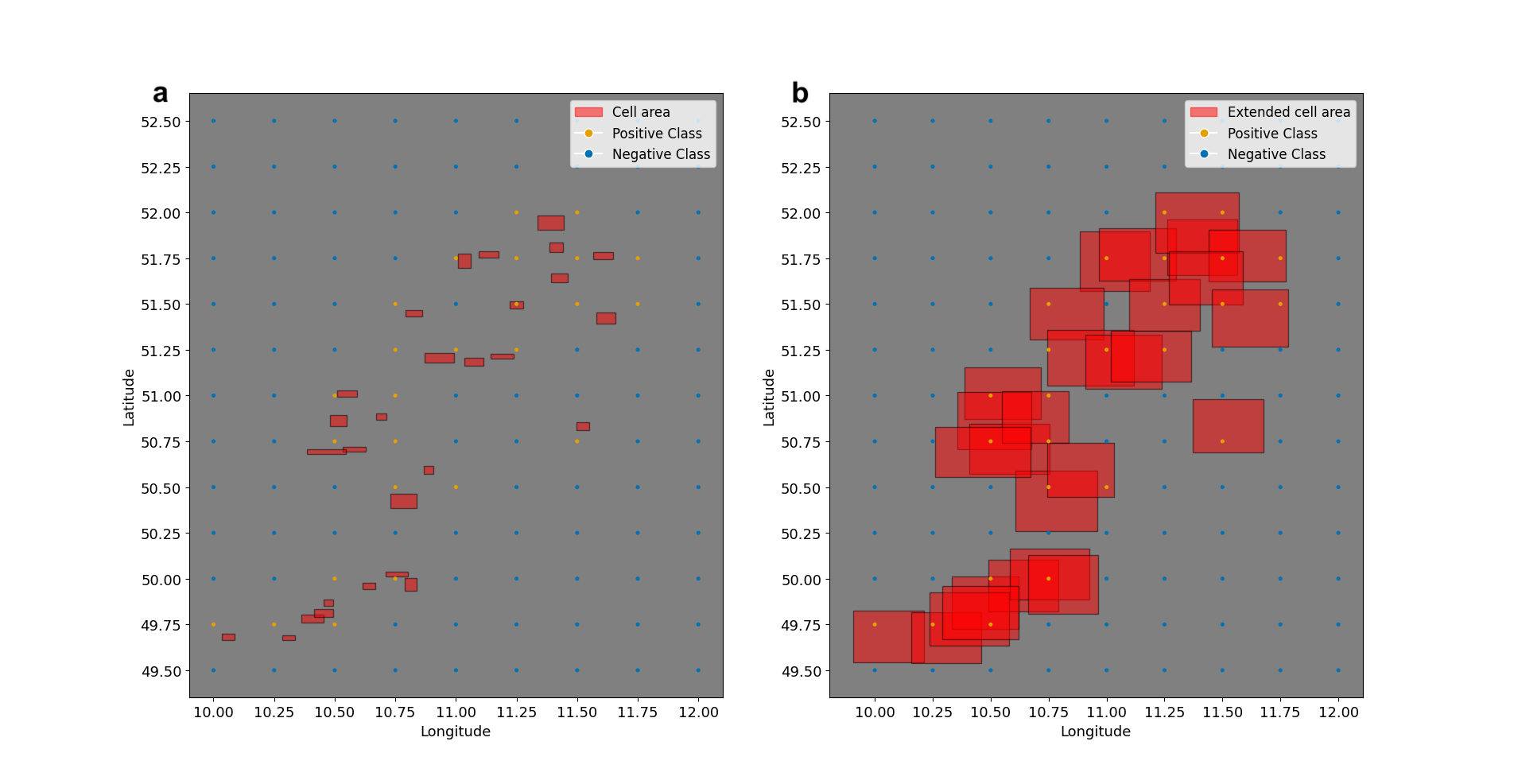}
    \caption{Example region in Germany on 10 July 2014 19 UTC showing how grid points are labelled the positive or negative class. Figure \textbf{a} shows the original cell area and \textbf{b} shows the cell areas extended by 0.125 degrees. Grid points falling within the extended cell area bounds are labelled the positive class. Positive and negative grid points are coloured orange and blue, respectively.}
    \label{fig:labelling}
\end{figure}

\begin{figure}
    \centering
    \includegraphics[trim={0cm 0cm 0cm 0cm},clip,scale=0.5]{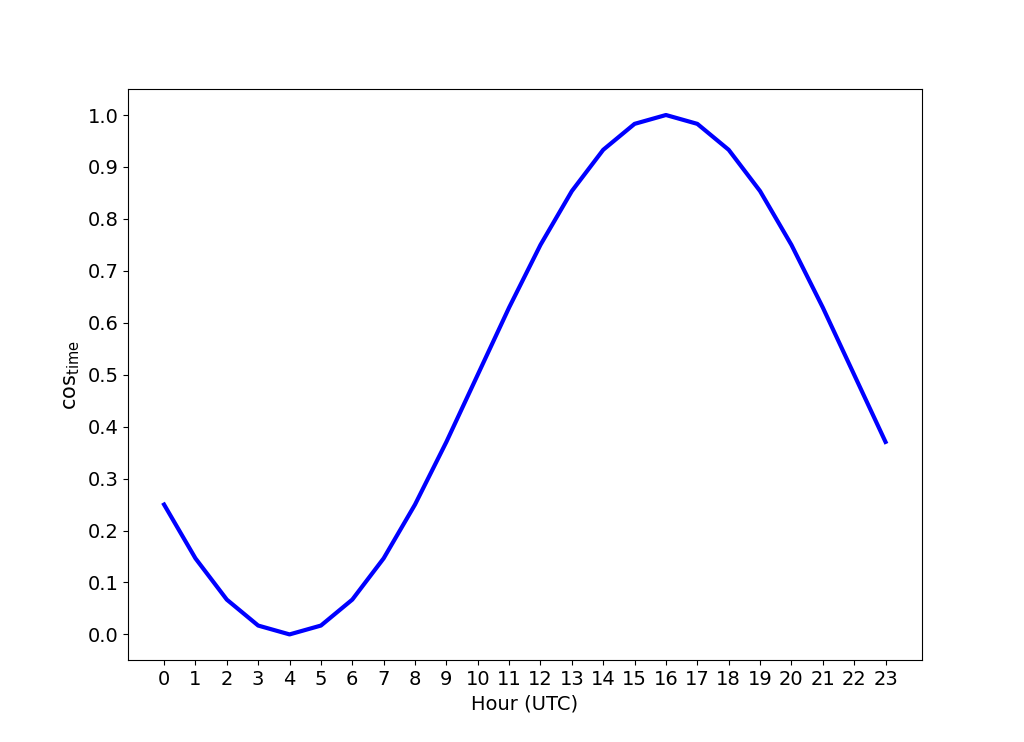}
    \caption{cos$_{time}$ as a function of Hour (UTC)}
    \label{fig:cosine_time_function}
\end{figure}

\begin{figure}
    \centering
    \includegraphics[trim={0cm 0cm 0cm 0cm},clip,scale=0.5]{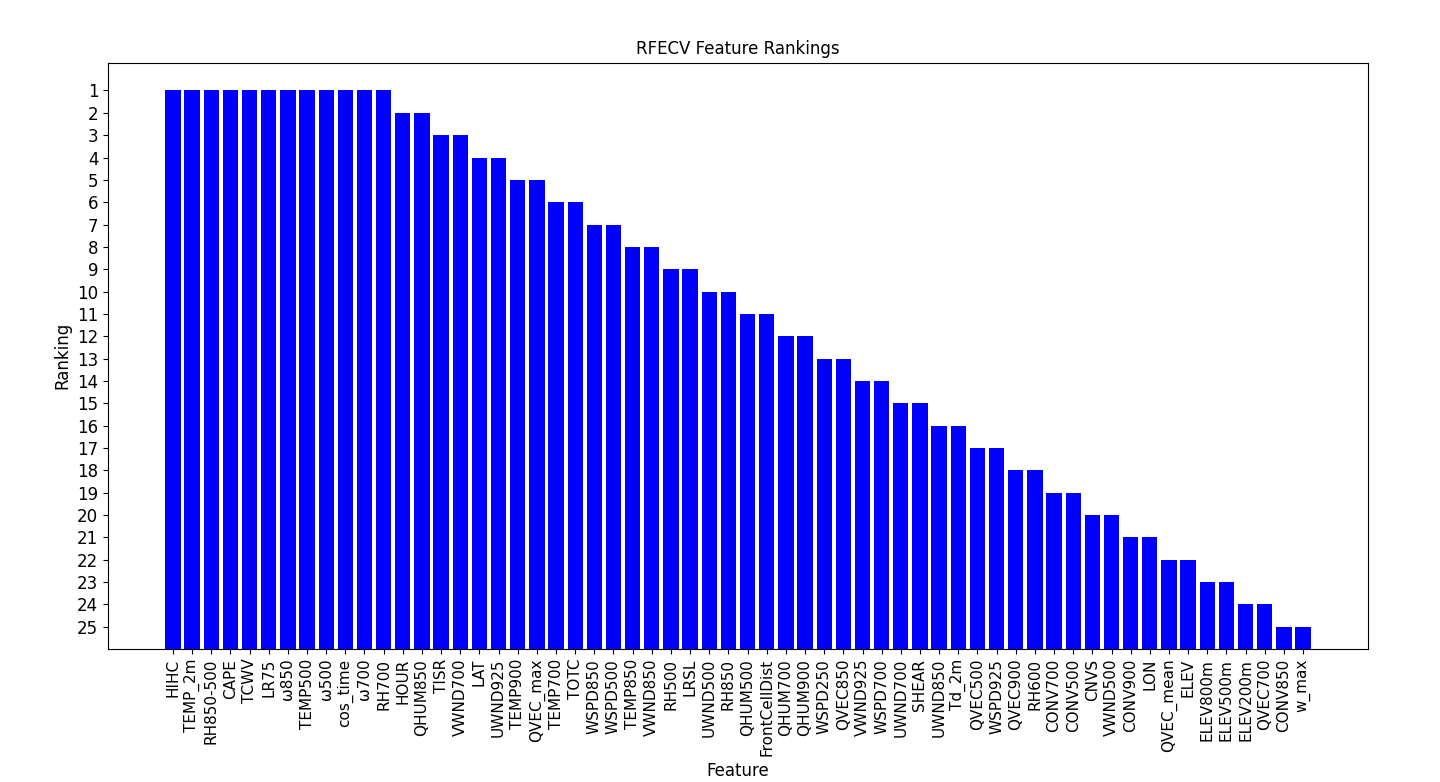}
    \caption{Feature ranking from 1 to 25 where lower numbers (left) indicate higher ranking and importance. Features with the same ranking are not ordered on the x-axis in terms of importance. Features with ranking number 1 are used in the final model and their importance is shown in Figure \ref{fig:best12_feature_importance}.}
    \label{fig:RFECV_ranking}
\end{figure}

\begin{figure}
    \centering
    \includegraphics[trim={1cm 3cm 1cm 3cm},clip,scale=0.5]{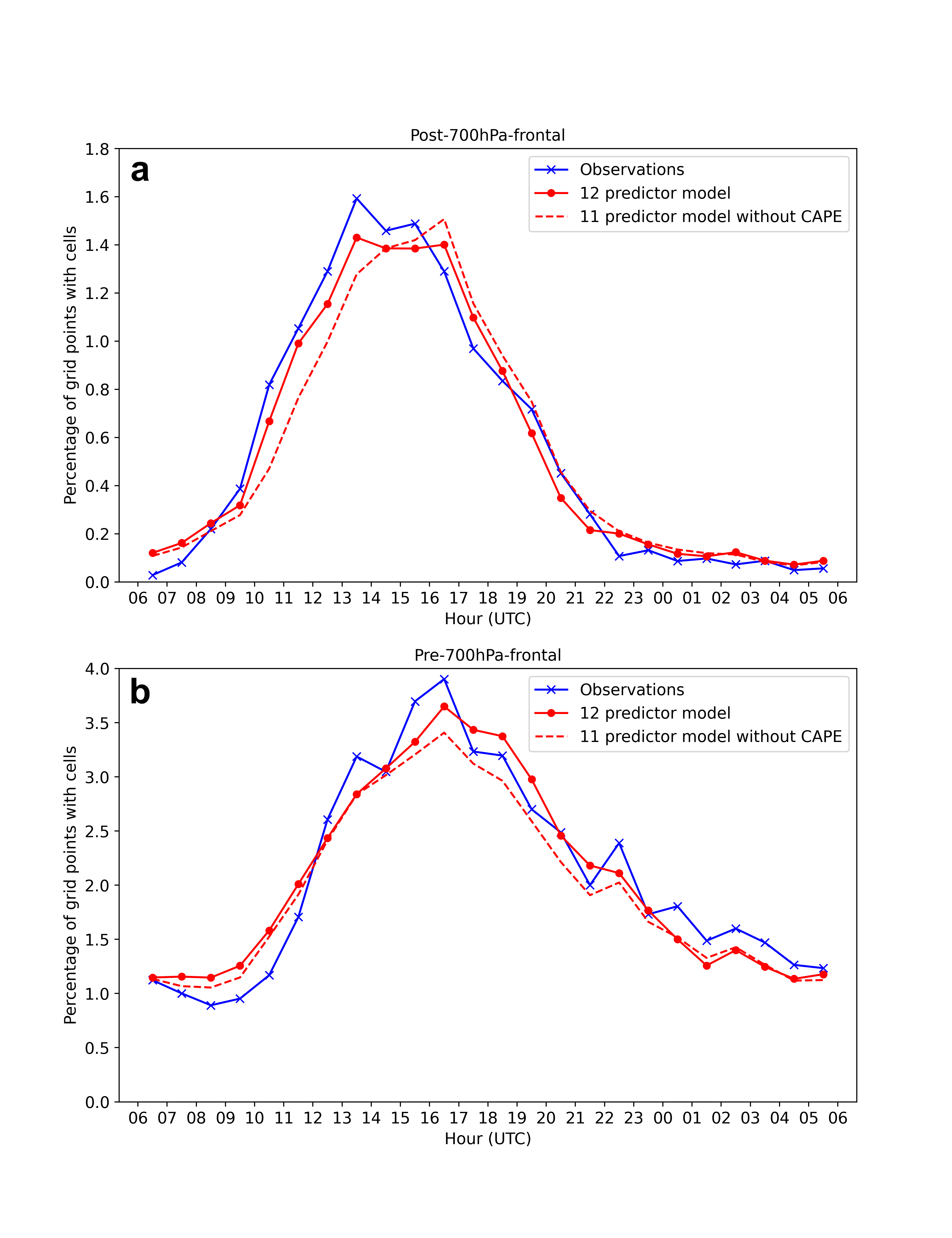}
    \caption{As Figure \ref{fig:diurnal_cycle_pre_post} but with CAPE omitted from the 11 predictor model instead of cosine time of day. }
    \label{fig:diurnal_cycle_pre_post_without_CAPE}
\end{figure}

\begin{figure}
    \centering
    \includegraphics[trim={0cm 0cm 0cm 0cm},clip,scale=0.5]{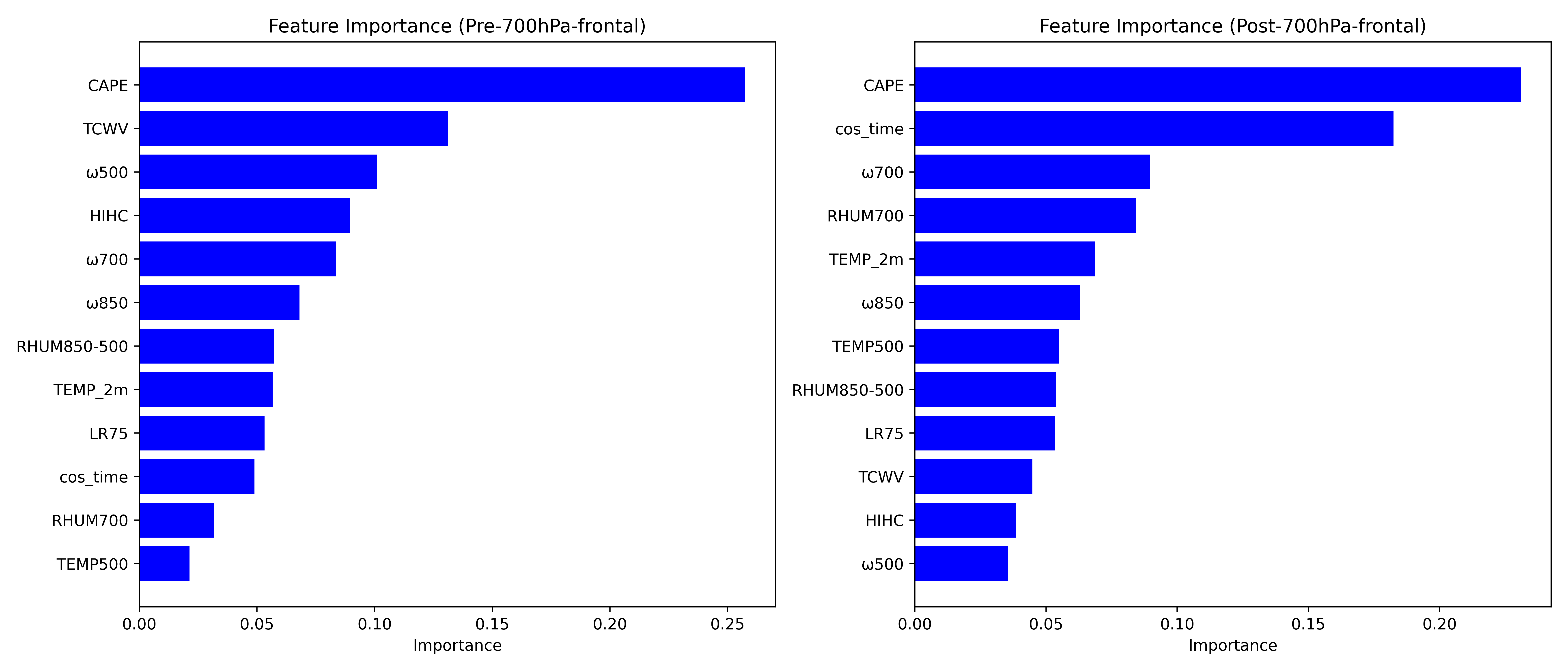}
    \caption{As Figure \ref{fig:best12_feature_importance} but training one model with pre-700hPa-frontal data and another model with post-700hPa-frontal data and then calculating the Gini impurity for each model.}
    \label{fig:two_models_feature_importance}
\end{figure}

\clearpage
\begin{table}[H]
    \centering
    \caption{Full predictor list with short name, units and levels. CAPE is derived considering parcels departing from different model levels below the 350 hPa level, and the departure level with the highest CAPE is retained (most unstable CAPE; MUCAPE). }
    \begin{tabular}{|l|l|l|l|}
        \hline
        \textbf{Predictor} & \textbf{Short Name} & \textbf{Units} & \textbf{Level} \\ \hline
        Dewpoint & Td$_{2m}$ & °C &  2-metres AGL\\ \hline
        Specific humidity & QHUM & kg kg\textsuperscript{-1} & 900, 850, 700, 500 \\ \hline
        Total column water vapour & TCWV & kg m\textsuperscript{-2}& integrated through entire atmosphere\\ \hline
        Relative Humidity & RH & \% & 850, 700, 600, 500, 850-500hPa average \\ \hline
        Convective Available Potential Energy & CAPE (MUCAPE) & J kg\textsuperscript{-1} & integrated between LFC and EL \\ \hline
        Maximum theoretical updraught velocity & w$_{max}$ & m s \textsuperscript{-1} & integrated between LFC and EL \\ \hline
        Lapse Rates & LR75, LRSL & °C & 700–500 hPa, surface–850 hPa \\ \hline
        Air temperature & TEMP & °C & surface, 900, 850, 700, 500 hPa \\ \hline
        Total Incoming Solar Radition & TISR & W m\textsuperscript{-2} & surface \\ \hline
        Q-vector convergence & QVEC & m\textsuperscript{2} kg\textsuperscript{-1} s\textsuperscript{-1}& 900,850,700,500,900-500max,900-500mean \\ \hline
        Convergence of wind fields & CONV & s$^{-1}$ & 900, 850, 700, 500 hPa \\ \hline
        Vertical Velocity & $w$ & m s\textsuperscript{-1} & 850, 700, 500 hPa \\ \hline
        Wind shear & WSHR & m s\textsuperscript{-1} & surface–500 hPa \\ \hline
        U component of wind & UWND & m s\textsuperscript{-1} & 925, 850, 700, 500 hPa \\ \hline
        V component of wind & VWND & m s\textsuperscript{-1} & 925,850,700,500 hPa \\ \hline
        Wind speed & WSPD & m s\textsuperscript{-1} & 925, 850, 700, 500, 250 hPa \\ \hline
        Total Cloud Cover & TOTC & \% & single level \\ \hline
        High Cloud Cover & HIHC & \% & single level (cloud above 6300 metres) \\ \hline
        Elevation & ELEV & km & single level \\ \hline
        Distance from elevation & ELEV(height)m & km & 200m, 500m, 800m \\ \hline
        Hour of the day & HOUR & hours & single level \\ \hline
        Cosine transformation of time of day & cos$_{time}$ & dimensionless & single level \\ \hline
        Distance from the 700 hPa front & FrontCellDist & km & single level \\ \hline
        Longitude & LON & degrees & single level \\ \hline
        Latitude & LAT & degrees & single level \\ \hline
    \end{tabular}
    \label{table:full_list_of_predictors}
\end{table}

\clearpage

\section{Recursive feature elimination with cross-validation (RFECV)}\label{subsec:RFECV}
\addcontentsline{toc}{subsection}{Recursive feature elimination with cross-validation}

Recursive feature elimination (RFE) is a machine learning method for feature selection. A model is first trained with all available features. Features are then ranked based on their importance or contribution to the model's performance. RFE eliminates the least important feature(s) from the dataset. The process is repeated iteratively until the predefined number of features is reached or until a specific performance criterion is met. RFECV (RFE with cross-validation) performs in a similar way except that at each iteration of feature elimination, cross-validation is used to evaluate the model's performance. This means the feature importance is assessed using different training and testing splits and averaged over all folds. The RFECV process was carried out using the Python \textit{scikit-learn} package (\citeauthor{scikit-learn}, \citeyear{scikit-learn}).

\section{Hyperparameter Tuning and Class Imbalance Tests}\label{subsec:hyperparameter}


The number of boosting rounds, the learning rate, the maximum depth of trees, the sample size and the fraction of predictors used for each tree are tuned (\citeauthor{Friedman_2001}, \citeyear{Friedman_2001}). 
The optimal hyperparameters are chosen using a Bayesian Search with cross-validation (BayesCV) from the \textit{scikit-optimize} Python package (\citeauthor{Head_et_al_2024}, \citeyear{Head_et_al_2024}). A probabilistic model of the search space is first constructed to identify the most promising regions. In this way, the search can then be explored more efficiently and the optimal hyperparameters can be selected with minimal iterations. The selected hyperparameters are shown below:

\begin{align*}
\text{number of boosting rounds} & : 558 \\
\text{maximum tree depth} & : 3 \\
\text{learning rate} & : 0.05 \\
\text{fraction of features used for each tree} & : 0.5 \\
\text{subsample size} & : 0.5 \\
\end{align*}

\noindent The hyperparameter 'scale\_pos\_weight', which adds an additional penalty to the minority class in the loss function, was also tested but did not add any improvement to the skill scores. Furthermore, keeping scale\_pos\_weight as 1 (no penalty) and using the hyperparameters above, the majority (negative) class was randomly undersampled such that the training data was more balanced. Training datasets with the following positive class percentages were produced by randomly undersampling the majority class: 1.5\%, 2.0\%, 2.5\%, 3.0\%, 4.0\%, 5.0\%, 6.0\%, 8.0\%, 10.0\%, 12.5\%, 15\%, 20\%, 25\%, 30\%, 40\%, 50\%. The testing dataset remained constant. The undersampling class imbalance tests also did not improve the skill score metrics.

\bibliographystyle{abbrvnat}
\bibliography{references}  


\end{document}